\documentclass[aps,prd,preprint,floatfix]{revtex4}
\usepackage{epsfig}
\usepackage{bm}

\begin{document}
  
\preprint{\rightline{ANL-HEP-PR-06-38}}
  
\title{Evidence for O(2) universality at the finite temperature transition 
for lattice QCD with 2 flavours of massless staggered quarks}
  
\author{J.~B.~Kogut}\thanks{Supported in part by NSF grant NSF PHY03-04252.}
\affiliation{Department of Energy, Division of High Energy Physics, Washington,
DC 20585, USA}
 \author{\vspace{-0.2in}{\it and}}
\affiliation{Dept. of Physics -- TQHN, Univ. of Maryland, 82 Regents Dr., 
College Park, MD 20742, USA}
\author{D.~K.~Sinclair}\thanks{This work was supported by the U.S.
Department of Energy, Division of High Energy Physics, Contract \\*[-0.1in]
W-31-109-ENG-38.}
\affiliation{HEP Division, Argonne National Laboratory, 9700 South Cass Avenue,
Argonne, IL 60439, USA}
                                                                                
\begin{abstract}
We simulate lattice QCD with 2 flavours of massless quarks on lattices of
temporal extent $N_t=8$, to study the finite temperature transition from
hadronic matter to a quark-gluon plasma. A modified action which incorporates
an irrelevant chiral 4-fermion interaction is used, which allows simulations at
zero quark mass. We obtain excellent fits of the chiral condensates to the
magnetizations of a 3-dimensional $O(2)$ spin model on lattices small enough
to model the finite size effects. This gives predictions for correlation 
lengths and chiral susceptibilities from the corresponding spin-model 
quantities. These are in good agreement with our measurements over the 
relevant range of parameters. Binder cumulants are measured, but the errors
are too large to draw definite conclusions. From the properties of the $O(2)$
spin model on the relatively small lattices with which we fit our `data', we
can see why earlier attempts to fit staggered lattice data to leading-order
infinite-volume scaling functions, as well as finite size scaling studies, 
failed and led to erroneous conclusions.
\end{abstract}

\maketitle

\section{Introduction}

When hadronic matter is heated to temperatures of 150--200~MeV it makes a
transition to a quark-gluon plasma. Such temperatures were present in the
early universe. There is also evidence that temperatures this high can be
observed in relativistic heavy-ion colliders such as RHIC. Indeed, claims that
the quark-gluon plasma has been observed have been made at RHIC and at CERN.

Lattice QCD enables one to simulate QCD at finite temperature and to examine
the nature of this transition. When hadronic matter passes through this
transition to the quark-gluon plasma, quarks and gluons are no longer confined.
In addition, the approximate chiral symmetry associated with light quarks,
which is broken spontaneously in cold hadronic matter, is restored. If we
reduce the quark mass to zero, this transition must necessarily become a phase
transition. We will consider the case of 2 light quarks ($u$ and $d$), and
take their masses to zero, assuming that the strange quark, which we neglect,
is too heavy to induce a first-order transition. For 2 light quark flavours,
it is believed that the zero-mass transition is a second-order phase
transition, which becomes a crossover with no true phase transition as soon as
the quarks gain mass \cite{Pisarski:1983ms}. It is argued that the
universality class of this phase transition is that of a 3-dimensional $O(4)$
spin model \cite{Pisarski:1983ms}. (Since the symmetry of massless 2-flavour
QCD is $SU(2) \times SU(2)$, which is isomorphic to $O(4)$, the
$\pi$--$\sigma$ multiplet which provides the Goldstone pions, expresses $O(4)$
symmetry.) The reduced symmetry of staggered quarks reduces this symmetry to
$O(2)$, so we might expect that lattice QCD with 2 staggered quark
flavours has a phase transition in the universality class of the 3-dimensional
$O(2)$ spin model.

In the standard formulations of lattice QCD, the Dirac operator becomes
singular at zero quark mass, and so zero-mass simulations are impossible and
very light quark simulations become very expensive. We have introduced a
modified version of the staggered quark action, which has an additional
irrelevant chiral 4-fermion interaction, ``$\chi$QCD'' \cite{Kogut:1998rg}. 
Being irrelevant, this 4-fermion interaction does not affect the continuum
limit, but it does render the Dirac operator non-singular, even in the chiral
limit. Hence we simulate 2-flavour QCD with massless quarks using the
$\chi$QCD action.

We present the results of simulations of massless two-flavour $\chi$QCD on
$N_t=8$ ($12^3 \times 8$, $16^3 \times 8$ and $24^3 \times 8$) lattices, in the
neighbourhood of the finite-temperature transition. State-of-the-art
simulations with conventional and highly improved staggered-quark actions have
been run at masses which are too large to accurately determine the
universality class of the transition \cite{Bernard:1996zw,Bernard:1997an}. As
a consequence, attempts to determine the universality class of this
transition, or to fit its scaling properties under the assumption that they
are those for $O(4)$ or $O(2)$ have produced confusing results
\cite{Karsch:1993tv,Karsch:1994hm,Aoki:1998wg,Bernard:1996cs,Bernard:1999fv,
Boyd:1996sk,Laermann:1998pf,Engels:2001bq}. 

Our earlier work on $N_t=4$ \cite{Kogut:1998rg,Kogut:2002rw} and $N_t=6$
\cite{Kogut:2000qp,Kogut:2001qu} lattices had produced results which appeared
to preclude either $O(2)$ or $O(4)$ universality. We had hoped that running at
$N_t=8$ would broaden the transition sufficiently to allow a better
determination of its nature. However, preliminary analysis
of our measurements indicated that the chiral condensate decreases too fast
for an $O(2)$ or $O(4)$ interpretation. The major problem appears to be that
our `data' needs to be extrapolated to infinite volume, and all such
extrapolations became unreliable close to the transition. For this reason we
abandoned our attempts to extrapolate to infinite volume, and instead fit our
chiral condensates to the magnetization curves for the 3-dimensional $O(2)$
spin model, also on finite volumes. By choosing the lattice volumes for best
fits, we are able to find good fits to our chiral condensates over most of the
range of our measurements. We are also able to use our spin model fits to make
acceptable predictions for the correlation lengths and chiral susceptibilities.

Once we establish that the behaviour of lattice QCD appears to be consistent
with that of the $O(2)$ spin model, we are able to use the properties of this
simple model obtained both from our own simulations and the extensive 
literature on this model \cite{Engels:2000xw,Cucchieri:2002hu,
Hasenbusch:1999cc,Ballesteros:1996bd,Engels:2001bq,Tominaga:1994bc} to
understand why naive attempts to establish $O(2)$ universality failed. The
sizes of the $O(2)$ lattices that give good fits to our $\chi$QCD `data' are
too small ($5^3$--$9^3$) to be anywhere close to the infinite volume limit. In
addition, the region over which the leading term suffices to describe the
critical scaling of the magnetization of the $O(2)$ spin model in the infinite
volume limit is very narrow \cite{Engels:2000xw}. A finite size scaling
analysis of the $O(2)$ spin model's magnetization reveals how easy it is to
misidentify the universality class of the transition, when only small-lattice
data is available.

In section~2 we introduce $\chi$QCD. Section~3 introduces the $O(2)$ spin model
and its critical behaviour. We describe our $N_t=8$ $\chi$QCD simulations and
present some of our results in section~4. In section~5 we present our analysis
of the $\chi$QCD `data' using the $O(2)$ spin model. Section~6 is devoted to
discussions and conclusions.

\section{``$\chi$QCD''}

In $\chi$QCD, the lattice Dirac operator is
\begin{equation}
A = \not\!\! D + m + \frac{1}{16} \sum_i (\sigma_i+i\epsilon\pi_i)
\end{equation}
where the first 2 terms are the standard staggered quark Dirac operator for
quarks in the fundamental representation of colour $SU(3)$, in interaction with
the gauge fields on the links of the lattice. The auxiliary fields $\sigma$
and $\pi$ are defined on the sites of the dual lattice and the sum is over the
16 sites of the dual lattice closest to the site on which the quark field
resides. $\epsilon(x)=(-1)^{x+y+z+t}$. The action for these auxiliary fields
is
\begin{equation}
S_{\sigma\pi} = \sum_{\tilde{s}}\frac{1}{8}N_f\gamma(\sigma^2+\pi^2)
\end{equation}
where the sum is over the sites of the dual lattice. Since this action is
completely local, these auxiliary fields can be integrated out introducing a
4-fermion interaction with coupling inversely proportional to $\gamma$.
$\chi$QCD preserves the remnant $U(1)$ chiral symmetry of the standard 
staggered quark action for massless quarks. The gauge-field action is the
standard Wilson action.

We use hybrid molecular-dynamics simulations to incorporate the factor of
$[\det (A^{\dagger} A)]^{N_f/8}$ in our simulations. Note that, in contrast to
the standard staggered Dirac operator, $A^{\dagger} A$ mixes even and odd 
lattice sites so that we cannot take the square root of the determinant by
restricting our noisy fermion fields to even (or odd) sites.

\section{The $O(2)$ spin model in 3 dimensions and critical behaviour}

The simplest version of the $O(2)$ spin model in 3 dimensions is a non-linear
$O(2)$ sigma model. The Hamiltonian ${\cal H}$ (strictly ${\cal H}/T$ in
condensed matter physics or the action if we are studying this model as a
field theory) is given by
\begin{equation}
{\cal H} = -J \sum_{<i,j>}\bm{s}_i\bm{\cdot}\bm{s}_j 
           - \bm{H\cdot}\sum_i\bm{s}_i ,
\label{eqn:O2}
\end{equation}
where $i$ and $j$ label sites on a 3 dimensional cubic lattice and $<i,j>$
run over all pairs of nearest-neighbour sites. The spins $\bm{s}$ are 
2-vectors of unit length. $\bm{H}$ is an external, symmetry-breaking magnetic
field. It is useful to define a temperature $T$ by $T=1/J$. This model and
variants have been studied extensively in the literature. Papers we have found
especially useful are listed in references~\cite{Engels:2000xw,Cucchieri:2002hu,
Hasenbusch:1999cc,Ballesteros:1996bd,Engels:2001bq,Tominaga:1994bc}.

The magnetization $\bm{M}$ of this system is defined as the lattice average 
of the spins
\begin{equation}
\bm{M} = \frac{1}{V}\sum_{\bm{i}} \bm{s}_{\bm{i}} ,
\end{equation}
where $V$ is the volume of the lattice. We first consider the thermodynamic
limit $V \rightarrow \infty$. At $\bm{H}=\bm{0}$, in the low temperature phase,
$O(2)$ is broken by a spontaneous magnetization with its associated Goldstone
mode. At high temperatures the magnetization vanishes and $O(2)$ symmetry is
restored. The phase transition occurs at a critical point $T=T_c=1/J_c$ with
$J_c=0.454165(4)$~\cite{Ballesteros:1996bd}.

Now let us define those critical exponents and amplitudes which describe the
critical behaviour of this model and which are relevant to what we might hope
to measure. Note that these definitions are more general
than this specific model. One defines a reduced temperature $t$ by 
$t=(T-T_c)/T_c$. We first consider the case $\bm{H}=\bm{0}$. As 
$t \rightarrow 0-$ the magnetization vanishes as
\begin{equation}
M=B(-t)^{\beta_m}.
\label{eqn:beta_m}
\end{equation}
As $t \rightarrow 0+$, the correlation length
\begin{equation}
\xi=z^+t^{-\nu},
\end{equation}
and the susceptibility
\begin{equation}
\chi = C^+ t^{-\gamma_m}.
\end{equation}
Here the susceptibility $\chi$ is defined by 
\begin{equation}
\chi = V \langle M^2 \rangle.
\label{eqn:susc}
\end{equation}
The correlation length $\xi$ can be defined in a number of different ways. We
will use the second moment formula
\begin{equation}
\xi = \left[{\chi/F-1 \over 4 \sin^2(\pi/L)}\right]^\frac{1}{2}
\label{eqn:xi2}
\end{equation}
where $L$ is the length of the box and F is defined by
\begin{equation}
F=\frac{1}{V}\left\langle\left|\sum_{\bm{i}} \exp (i p_z z)\bm{s}_{\bm{i}}
\right|^2\right
  \rangle
\end{equation}
where $z$ is the third coordinate of the site $\bm{i}$ and $p_z=2\pi/L$. This
form for $\xi$ appears to have originated with reference~\cite{Cooper:1982nn}.
At small but non-zero H, the magnetization
\begin{equation}
M=d H^\frac{1}{\delta}.
\end{equation}
For the $O(2)$ spin model these critical exponents have (approximate) values
$\beta_m = 0.3490(6)$, $\nu = 0.6723(11)$, $\gamma_m \approx 1.32$
$\delta \approx 4.7798$, $\omega = 0.79(2)$ ($\omega$ is a critical exponent
parameterizing finite size corrections). 

More information can be obtained about critical points by measuring
fluctuation quantities (in addition to the susceptibility). One such quantity
is the fourth-order Binder cumulant
\begin{equation}
B_4={\langle M^4 \rangle \over \langle M^2 \rangle^2}.
\end{equation}
For $H=0$ on an infinite lattice, this quantity is $1$, the value for a
first-order transition, for $T < T_c$, and $2$, the value for a crossover, for
$T > T_c$. Right at $T_c$ it has been measured at $B_4=1.242(2)$ for the
$O(2)$ spin model~\cite{Cucchieri:2002hu}. On a finite lattice there is a
smooth crossover from $1$ to $2$ as $T$ is varied from small to large values,
crossing $T_c$. This crossover steepens as $V$ is increased. For large $V$s a
finite size scaling analysis indicates that these curves cross at points which
approach $[T_c,B_4(T_c)]$ as $V \rightarrow \infty$. Similarly curves of $\xi/L$
for finite $L$ (and hence $V$) cross at points which approach $T_c$ and a
universal value $[\xi/L](T_c) = 0.593(2)$, a value specific to $O(2)$.

Working at finite volume, where $\langle\bm{M}\rangle = \bm{0}$, it is
necessary to find an alternative definition of the ensemble-averaged 
magnetization. One which has proved useful is
\begin{equation}
M = \langle |\bm{M}| \rangle,
\end{equation}
which will approach the value of the spontaneous magnetization as 
$V \rightarrow \infty$. It is clear, however, that this quantity does not
vanish in the high temperature phase except in the infinite volume limit.
Hence, to test that it has the correct critical behaviour or to measure the
critical exponent $\beta_m$, it is necessary either to extrapolate to 
infinite volume, or to apply a finite size scaling analysis. Finite size
scaling for $M$ predicts that close to the critical point \cite{barber}
\begin{equation}
M(t,L) = L^{-\frac{\beta_m}{\nu}}[Q(tL^\frac{1}{\nu})
                                 + L^{-\omega}G(tL^\frac{1}{\nu}) + ...]
\label{eqn:fss}
\end{equation}
so that plotting $L^\frac{\beta_m}{\nu} M$ against $tL^\frac{1}{\nu}$ should
make the magnetization curves for different $L$ values coincide, provided $L$
is sufficiently large. 

For QCD, the chiral condensate plays the r\^{o}le of the magnetization. In the
2-dimensional space of the remnant $U(1)$ or $O(2)$ symmetry of staggered
fermions, the chiral condensate $(\bar{\psi}\psi,i\bar{\psi}\gamma_5\xi_5\psi)$
or $(\sigma,\pi)$ corresponds to the magnetization of the $O(2)$ spin model.

\section{$\chi$QCD simulations on $N_t=8$ lattices}

We have performed high statistics simulations of massless 2-flavour $\chi$QCD
on $16^3 \times 8$ and $24^3 \times 8$ lattices and lower statistics
simulations on $12^3 \times 8$ lattices, using the hybrid molecular-dynamics
algorithm with $dt=0.05$. Measurements are made every 2 1-time-unit
trajectories. For these simulations the parameter $\gamma$ was fixed at $10$.
At each of 6 values of $\beta=6/g^2$ close to the transition
($\beta=5.5250,5.5275,5.5300,5.5325, 5.5350,5.5375$) we ran for 100,000
trajectories on each of the two larger lattices. For
$\beta=5.5400,5.5450,5.5500$ we ran for 50,000 trajectories on each of the 2
larger lattices. Shorter runs were made for $\beta$ values outside of this
range.

\begin{figure}[htb]
\epsfxsize=6in
\centerline{\epsffile{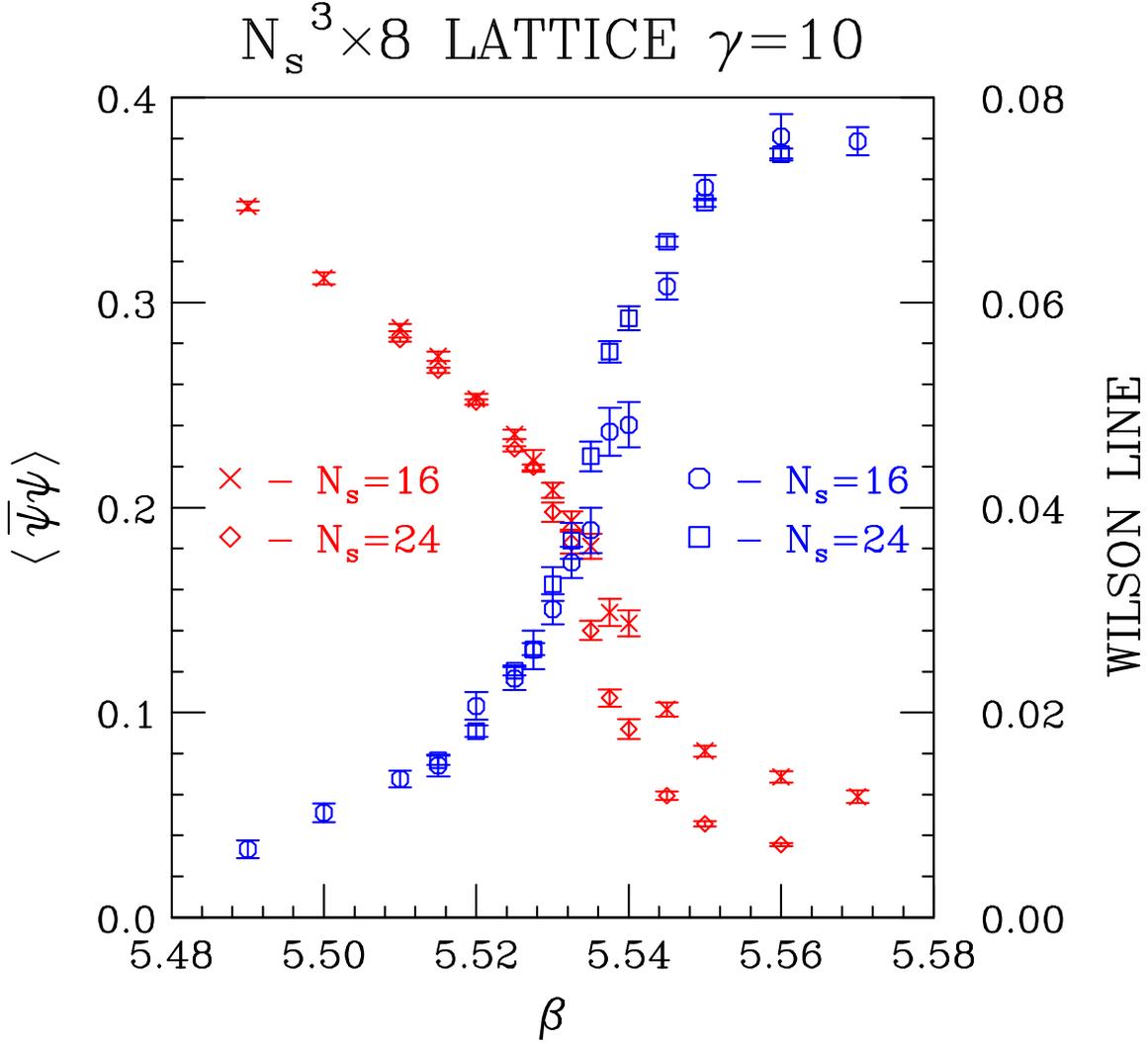}}
\caption{Chiral condensate and Wilson Line as functions of $\beta=6/g^2$ for
$16^3 \times 8$ and $24^3 \times 8$ lattices.}
\label{fig:wilpsi}
\end{figure}

\begin{figure}[htb]
\epsfxsize=4in
\centerline{\epsffile{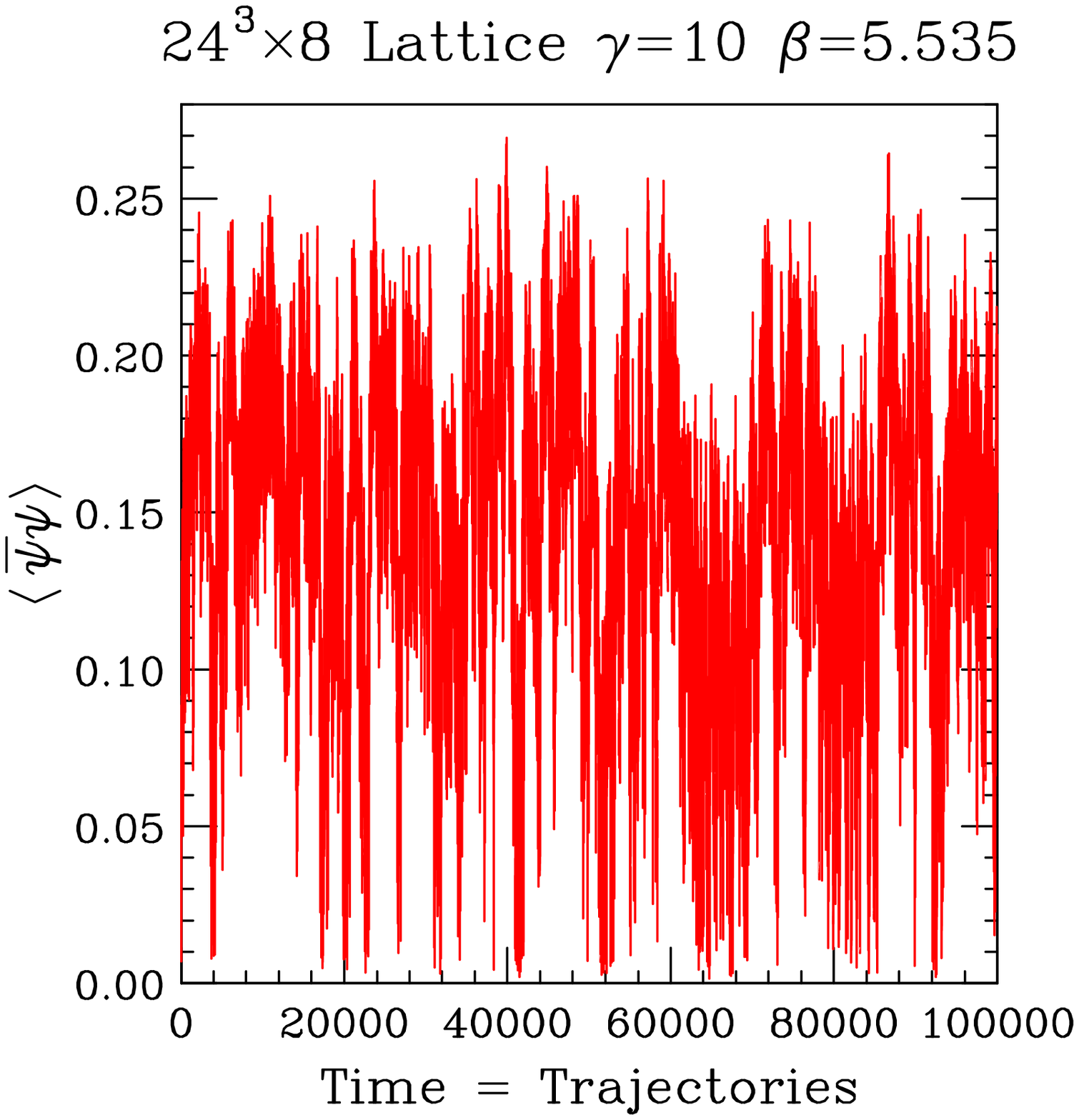}}
\vspace{0.25in}
\centerline{\epsffile{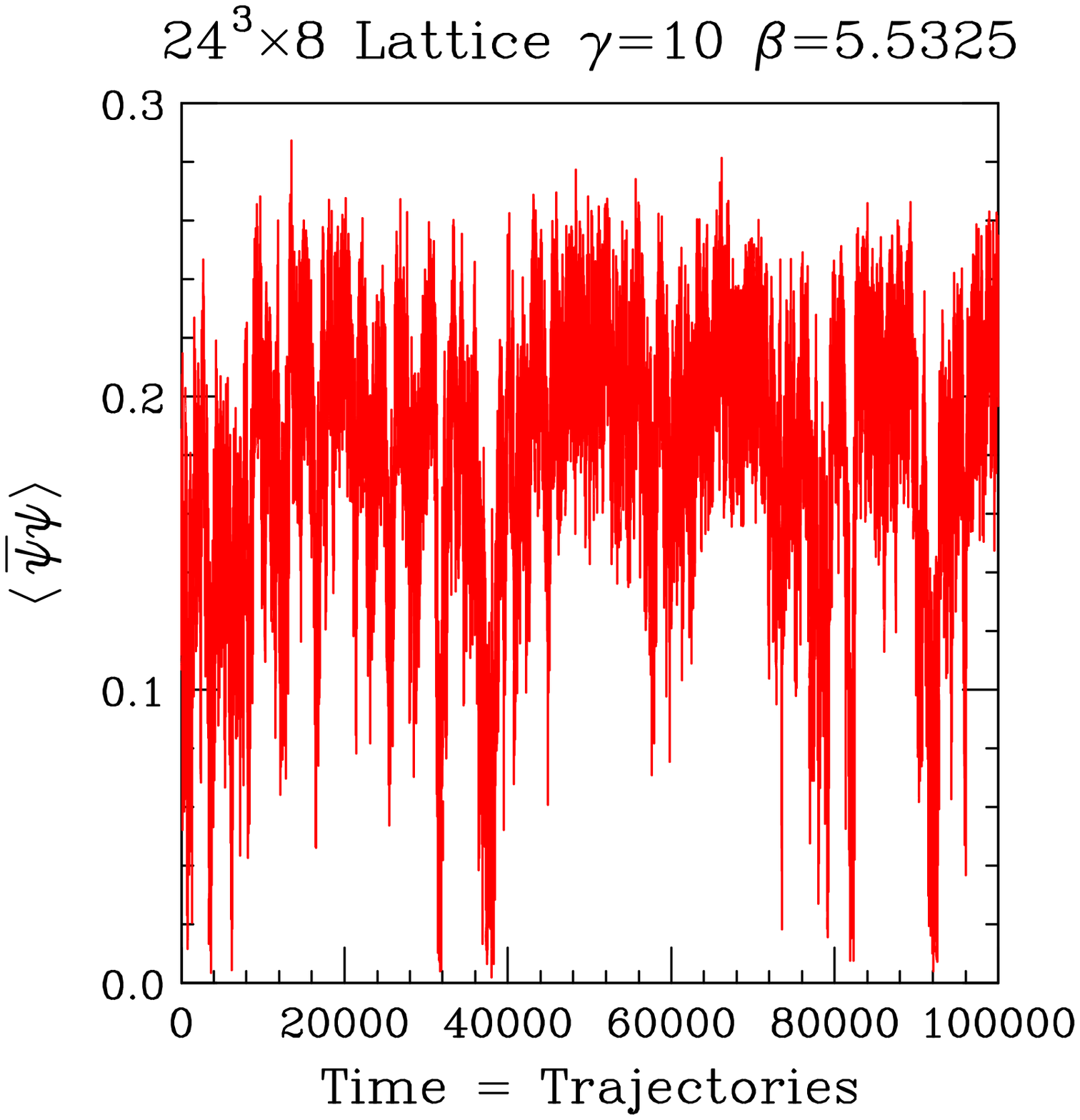}}
\caption{Time histories of the chiral condensate on a $24^3 \times 8$ lattice
a) for $\beta=5.535$ b) for $\beta=5.5325$.}
\label{fig:time}
\end{figure}

Figure~\ref{fig:wilpsi} shows the chiral condensate and Wilson line (Polyakov 
loop) as functions of $\beta=6/g^2$ for both the $16^3 \times 8$ and the
$24^3 \times 8$ lattices. From this figure, it is clear that the finite
temperature phase transition occurs somewhere in the range 
$5.53 \lesssim \beta \lesssim 5.54$. The smoothness of the transition suggests
strongly that it is second order. The curvature of the chiral condensate for
the larger lattice (before finite volume rounding sets in) indicates that
$\beta_m < 1$, something that the best finite mass simulations to date have
difficulty in observing. The reason that such high statistics are needed 
becomes clearer when one looks at the time histories of observables measured
during these runs. In figure~\ref{fig:time} we show the time history of the
chiral condensate on the $24^3 \times 8$ lattice for $\beta=5.535$ which is
close to the transition, and $\beta=5.5325$ which is just below the transition.
Because of critical slowing down, the system shows fluctuations which last for
thousands of trajectories close to the critical point.

\begin{figure}[h]
\epsfxsize=4.5in
\centerline{\epsffile{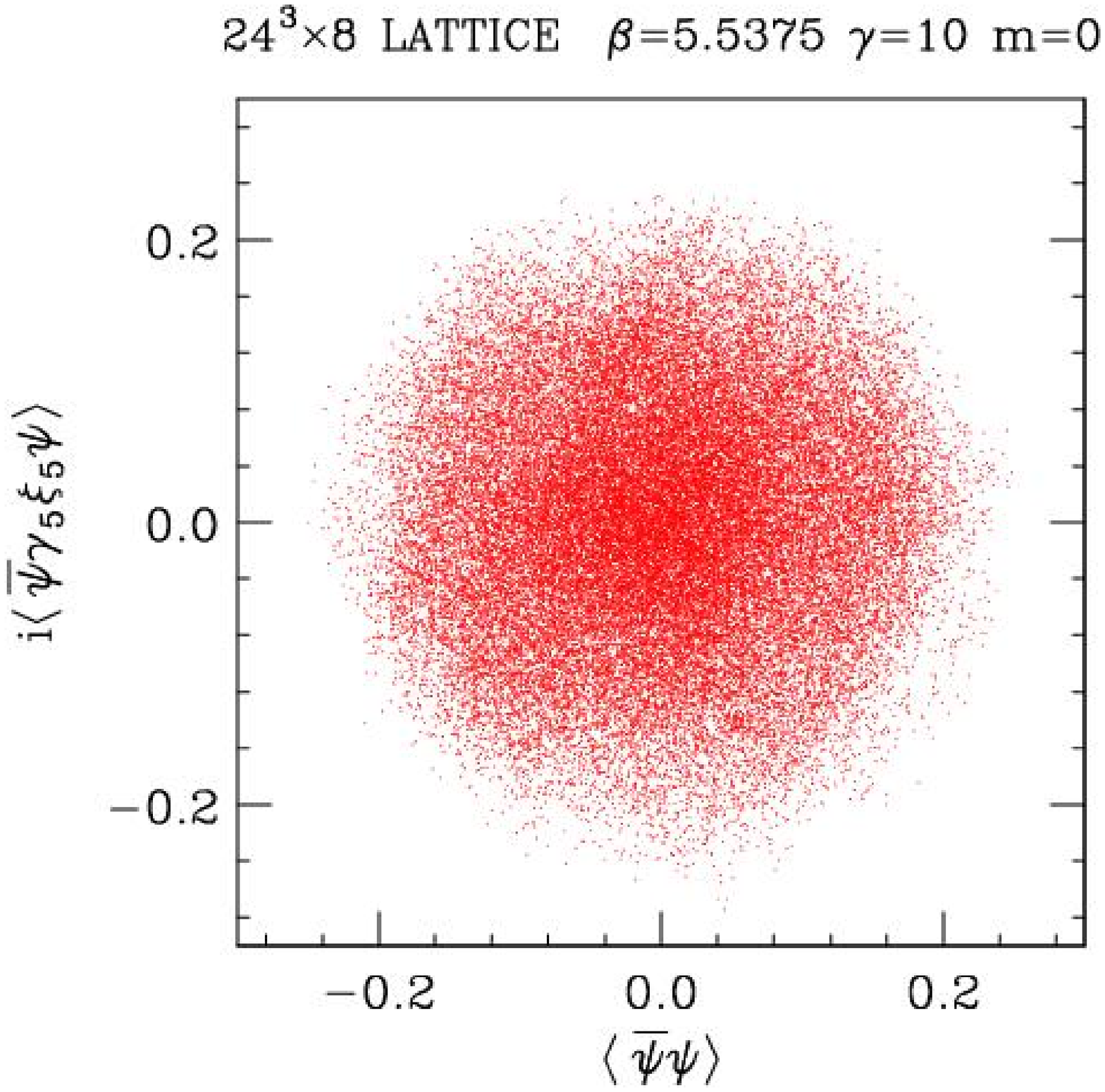}}
\caption{Scatter-plots of the chiral condensates on a $24^3 \times 8$ lattice.
Each of the 50,000 points on each graph represents the value on a single
configuration the angle brackets here indicate that these are configuration
averaged quantities but {\it not} ensemble averages. a) $\beta=5.5325$,
b) $\beta=5.5350$, c) $\beta=5.5375$.}
\label{fig:scatter}
\end{figure}

With the standard staggered action at zero quark mass, the chiral condensate
vanishes identically on each configuration, so even if we could simulate at
$m=0$ we could not observe a chiral condensate, even on a configuration by
configuration basis. This has to be true because the remnant $U(1)$ chiral
symmetry is manifest at any fixed value of the gauge fields, i.e. on each
configuration, which precludes the existence of a chiral condensate. With the
$\chi$QCD action, the condensate is finite on each configuration. This is
because, a configuration consists of a set of gauge fields {\it and} a set of
auxiliary fields $\sigma$ and $\pi$. Any non-zero set of auxiliary fields
breaks chiral symmetry. The vanishing of the chiral condensate on a finite
volume occurs because, as the system evolves in molecular-dynamics `time', the
condensate rotates in the $(\bar{\psi}\psi,i\bar{\psi}\gamma_5\xi_5\psi)$
plane so that the ensemble average is zero. This is because all $U(1)$ global
chiral rotations of the auxiliary fields contribute to the ensemble of
configurations with equal weight. What we have called
$\langle\bar{\psi}\psi\rangle$ in the figures above is really 
$\langle \sqrt{\bar{\psi}\psi^2-\bar{\psi}\gamma_5\xi_5\psi^2}\rangle$ where 
the quantities under the square root are lattice averaged before being squared.
Similarly we define a quantity which we denote $\langle\sigma\rangle$ which is
really $\langle\sqrt{\sigma^2+\pi^2}\rangle$. In the infinite volume limit
each of these quantities will approach the condensate of interest. These
condensates are related by
\begin{equation}
\langle\bar{\psi}\psi\rangle = \gamma \langle\sigma\rangle
\end{equation}
at $dt=0$, (at least in the infinite volume limit) so either quantity can be
used as the chiral condensate. This will be important when we consider
fluctuation quantities, since the `$\bar{\psi}\psi$' we measure is only a
stochastic estimator of the condensate, while $\sigma$ is a true condensate.
Figure~\ref{fig:scatter} is a scatter-plot of the actual chiral condensates
measured in the $(\bar{\psi}\psi,i\bar{\psi}\gamma_5\xi_5\psi)$ plane on
individual configurations for 3 $\beta$ values close to the transition, on a
$24^3 \times 8$ lattice. One should first note that the condensate does indeed
rotate in the plane between configurations so that the ensemble average is
zero, even in the ordered phase. At $\beta=5.5325$, below the transition, the
distribution is concentrated in a ring, indicating a non-zero condensate. By
$\beta=5.5350$, close to the transition, the ring structure is no longer clear,
although a peak away from the origin still shows up on a radial density
profile. Finally, for $\beta=5.5375$, above the transition, the maximum
density is at the origin indicating that the condensate has vanished, and
chiral symmetry is restored.

\begin{figure}[htb]
\epsfxsize=4.5in
\centerline{\epsffile{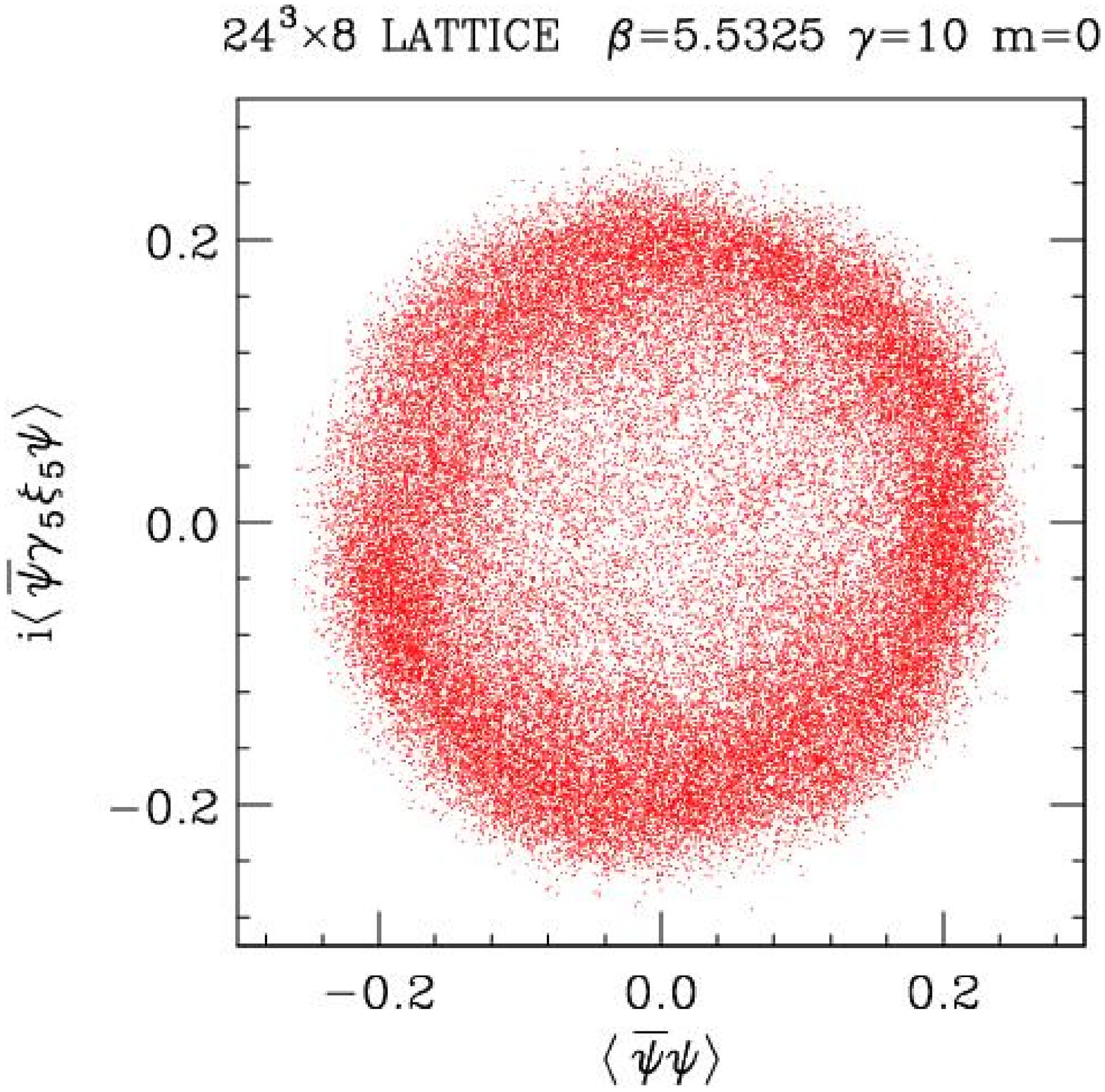}}
\vspace{0.25in}
\centerline{\epsffile{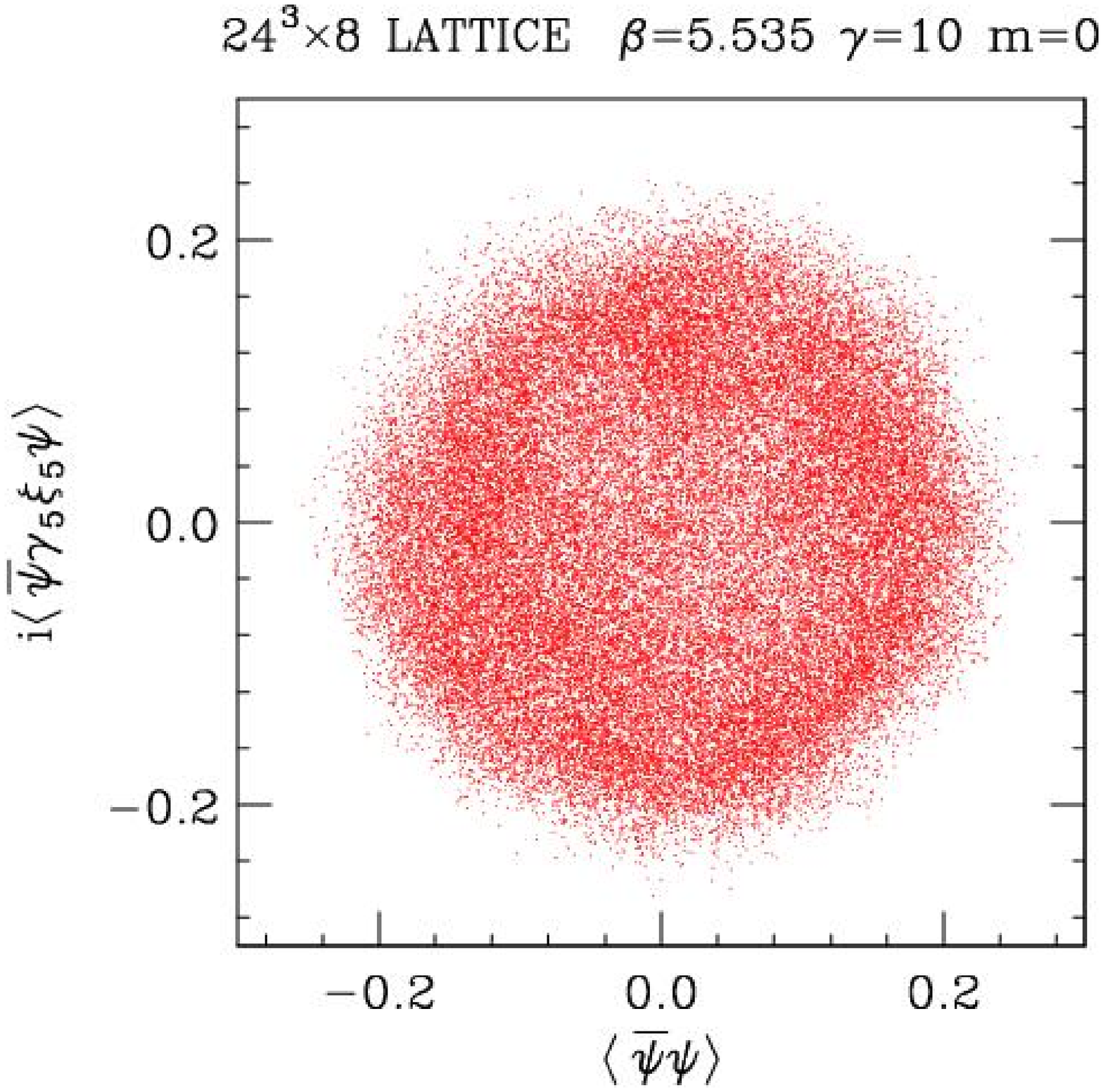}}
\end{figure}

In the figure~\ref{fig:wilpsi} it is clear that to attempt fits to the leading
critical scaling behaviour
\begin{equation}
\langle\bar{\psi}\psi\rangle = b (\beta_c-\beta)^{\beta_m} 
\;\;\;\;\; \beta < \beta_c,
\end{equation}
requires continuation to infinite volume. Unfortunately, such extrapolations
break down close to the transition, which is where accurate measurements are
most needed. Fits to any of these attempted extrapolations always favour 
values of $\beta_m$ much smaller than that of the 3-dimensional $O(2)$ spin
model. It is for this reason that we have tried fitting to the magnetization
curves for the $O(2)$ spin model on {\it small} lattices which also incorporate
finite volume effects.

\begin{figure}[htb]
\epsfxsize=6.0in
\centerline{\epsffile{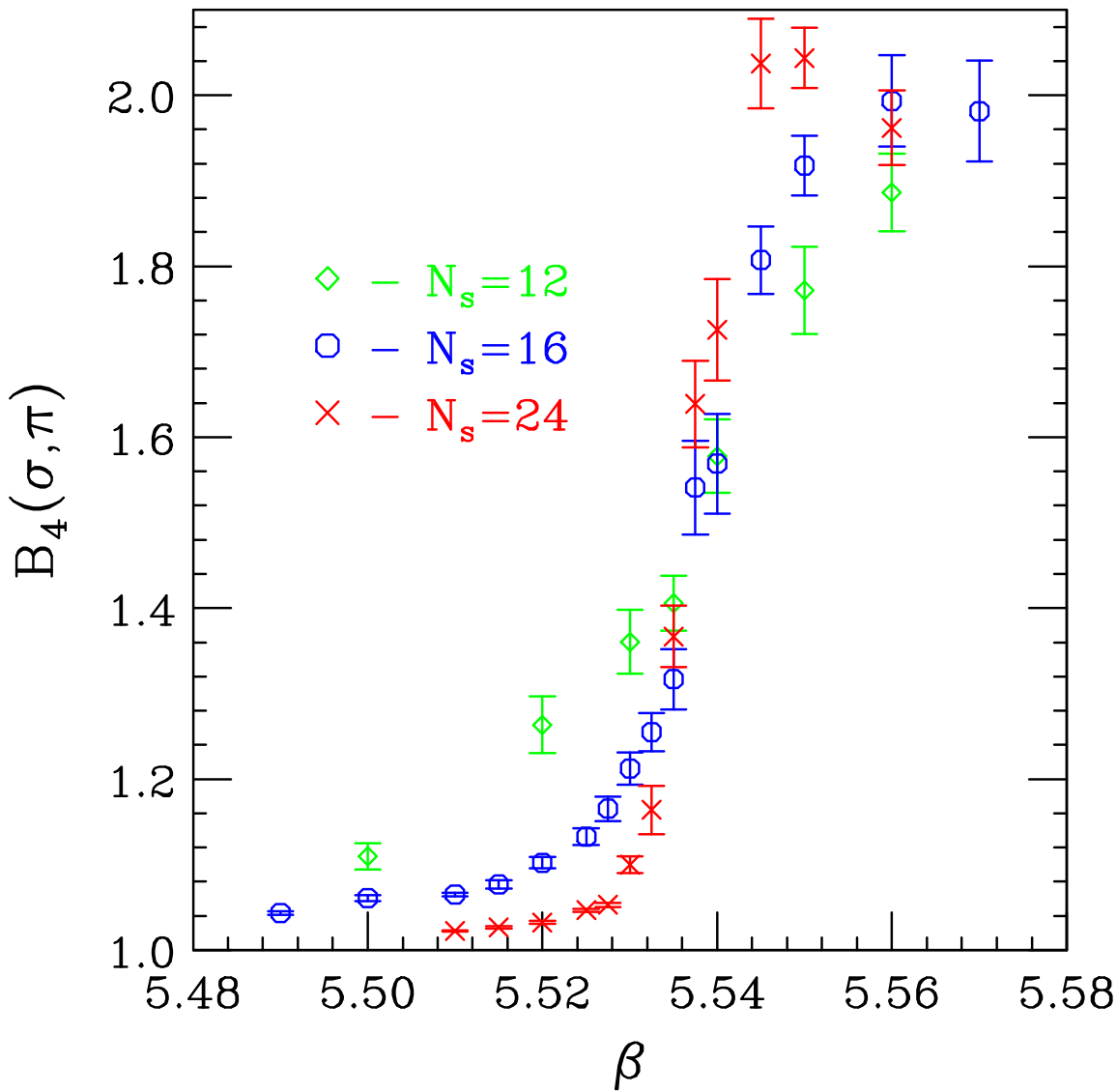}}
\caption{The fourth-order binder cumulants for the auxiliary fields as
functions of $\beta$ for $12^3 \times 8$, $16^3 \times 8$ and $24^3 \times 8$
lattices.}
\label{fig:binder}
\end{figure}

Binder cumulants \cite{Binder:1981sa} have been found useful in characterizing
phase transitions. Figure~\ref{fig:binder} shows the 4th order Binder
cumulants for the auxiliary fields $(\sigma,\pi)$. Following the definitions
given in the previous section, $B_4$ is defined by
\begin{equation}
B_4={\langle(\sigma^2+\pi^2)^2\rangle \over \langle\sigma^2+\pi^2\rangle^2}
\end{equation}
For low $\beta$, $B_4$ does appear to approach $1$ as expected for a first order
transition (this is the first order transition when the mass $m$ passes
through zero, causing the condensate to abruptly change sign). At high $\beta$,
$B_4$ does appear to approach $2$, the value for a crossover, as expected. In
between the curves for the 3 lattice sizes do appear to cross at an intermediate
value, which is characteristic of a second order transition. The statistics are
not good enough to determine the positions or values of these intersections,
but they appear to be higher than the value $1.242(2)$ expected if the 
transition were in the $O(2)$ universality class. However, this probably
indicates that our lattices are too small for this intersection to occur very
close to the infinite volume value. However, this powerful technique should be
an excellent way of determining the nature of the transitions for larger
lattices with very high statistics.

To get estimates of the position of the transition directly from the `data'
(rather than from the fits to spin-model curves) we look at the radial chiral
susceptibility, expressed in terms of the auxiliary fields
\begin{equation}
\chi_r(\bm{\sigma}) = V (\langle|\bm{\sigma}|^2\rangle 
                       - \langle|\bm{\sigma}|\rangle^2),
\end{equation}
where $\bm{\sigma}=(\sigma,\pi)$. The position of the phase transition is
estimated from the peak of this susceptibility, which is found using
Ferrenberg-Swendsen reweighting \cite{Ferrenberg:1988yz} to interpolate our
`data'. This yields an estimate of $\beta_c=5.5390(6)$ for the $16^3 \times 8$
lattice and $\beta_c=5.5359(5)$ for the $24^3 \times 8$ lattice. (Note that we
have also tried using Ferrenberg-Swendsen reweighting to interpolate the
`data' for our condensates. While this indicates that the interpolations from
the measurements at $\beta$ values at the ends of each interval are
consistent, the curves it yields at current statistics are not sufficiently
smooth to help with fits, taking into account that the extra points yielded by
these interpolations are highly correlated.) In the next section we indicate
how $\beta_c$ calculated in this way approaches the infinite volume limit. 

\section{$O(2)$ spin-model analysis of $\chi$QCD measurements}

\begin{figure}[htb]
\epsfxsize=3.5in
\centerline{\epsffile{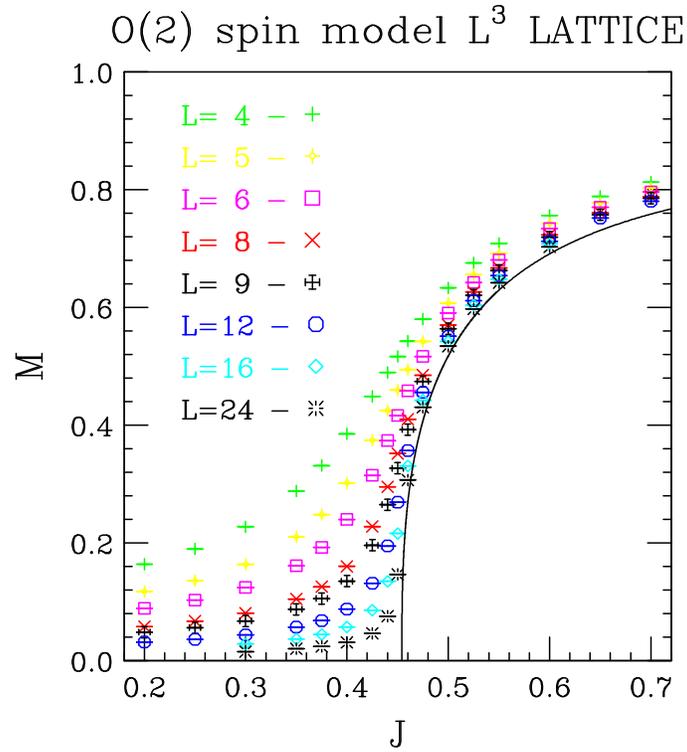}}
\vspace{0.25in}
\centerline{\epsffile{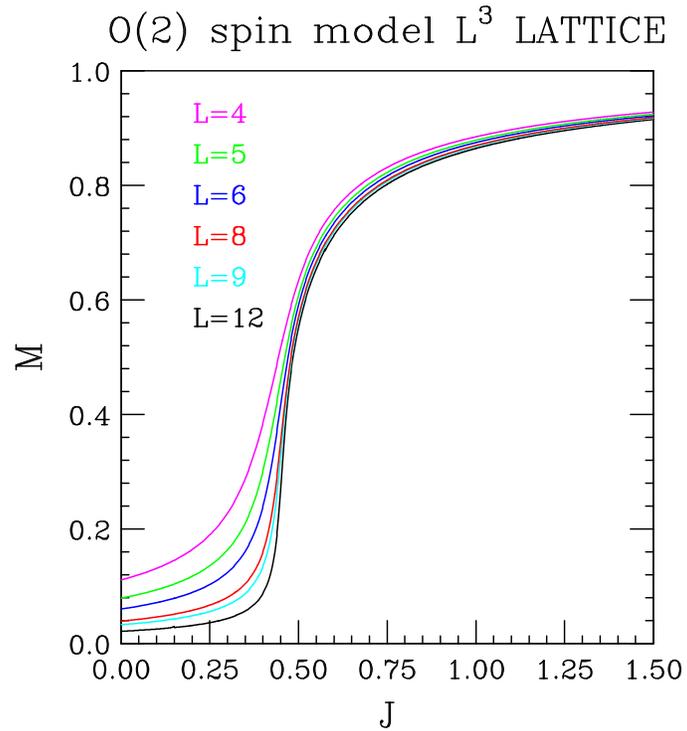}}
\caption{a) Magnetization as a function of coupling $J$ for the 3-dimensional
O(2) spin-model. The curve is a parametrization of the infinite volume limit
given in reference~\cite{Engels:2000xw}. b) Magnetization curves from 
Ferrenberg-Swendsen reweighting.}
\label{fig:mag}
\end{figure}

We have simulated the 3-dimensional $O(2)$ spin model with the Hamiltonian given
in equation~\ref{eqn:O2} on small lattices -- $4^3$, $5^3$, $6^3$, $8^3$, 
$9^3$, $12^3$, $16^3$ and $24^3$. These are adequate for the comparisons we
make with our results from lattice QCD simulations using the $\chi$QCD action,
which were presented in the previous section. For spin-model results on lattices
large enough to make comparison with the infinite volume limit, we appeal to
the literature. For our $4^3$ to $12^3$ lattices, we have interpolated
our measurements of the magnetization over the range $0 \le J \le 1.5$ using
Ferrenberg-Swendsen reweighting. Figure~\ref{fig:mag} shows the results of our
$O(2)$ simulations and the scaling curves they produced. 

Attempts to perform a finite size scaling analysis (equation~\ref{eqn:fss})
around the chiral phase transition of $\chi$QCD with $O(2)$ critical indices
fail to indicate the existence of a universal scaling function. Thus, if the
critical point of this theory is in the $O(2)$ universality class, finite
volume effects on lattices of the size we use are too large to allow the
`data' to be fit by a single scaling function. 

\begin{figure}
\epsfxsize=3.8in
\centerline{\epsffile{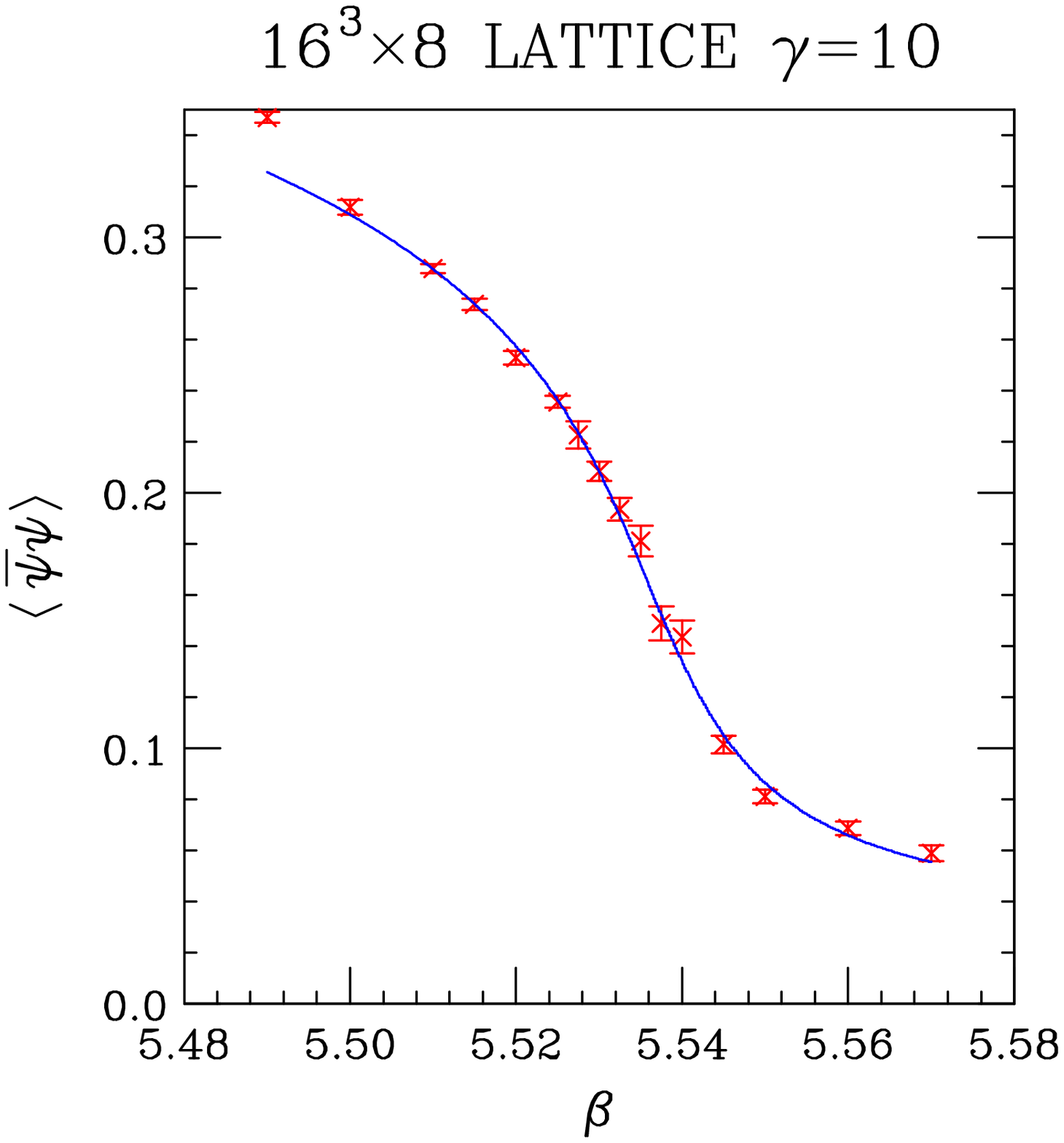}}
\vspace{0.25in}
\centerline{\epsffile{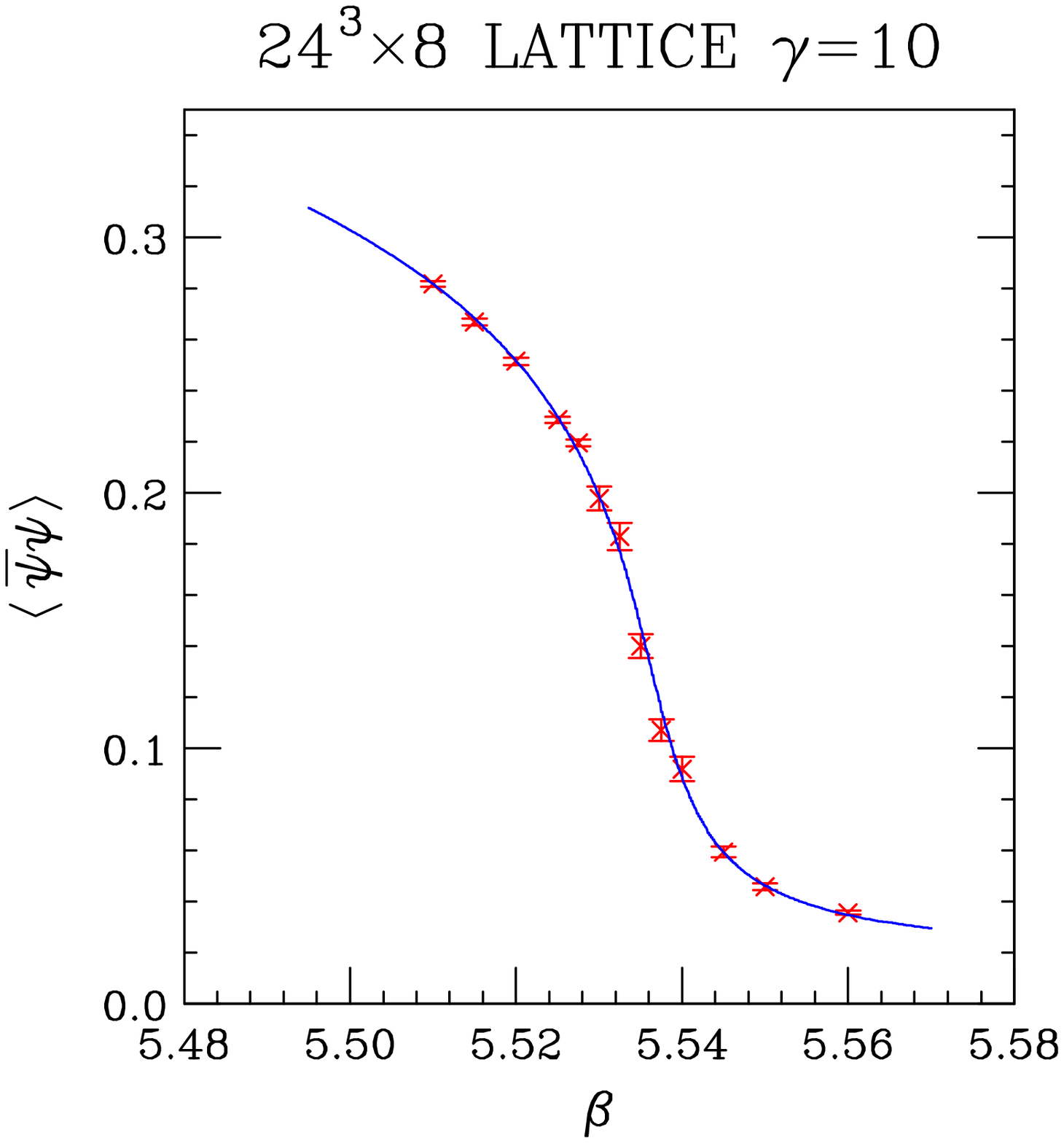}}
\caption{a) Fit of the chiral condensate for $\chi$QCD on a $16^3 \times 8$
lattice to the magnetization of an $O(2)$ spin model on a $6^3$ lattice.
b) Fit of the chiral condensate for $\chi$QCD on a $24^3 \times 8$
lattice to the magnetization of an $O(2)$ spin model on an $8^3$ lattice.}
\label{fig:condfits}
\end{figure}

An alternative approach, is to fit the chiral condensates to the magnetization
curves for the $O(2)$ spin models on lattices sufficiently small that they too
show finite size departures from the universal scaling function. 
Note that these departures depend on the specific form of the $O(2)$ spin-model 
Hamiltonian used (equation~\ref{eqn:O2}). 
We allow ourselves the freedom to vary the spin-model lattice sizes to obtain
the best fits to the $\chi$QCD `data'.
This is justified if a single operator is adequate to describe most of the
departures from the $L \rightarrow \infty$ universal scaling function of
equation~\ref{eqn:fss} for the $O(2)$ spin model, and the corresponding 
operator describes most of these departures for $\chi$QCD. Varying the lattice
size for the $O(2)$ spin model allows us to match the coefficients of the
universal scaling function (presumably the coefficient $G$ of $L^{-\omega}$
in equation~\ref{eqn:fss}) which describes this departure. If more than one
non-leading term is required for the model of equation~\ref{eqn:O2}, modeling 
non-leading behaviour of $\chi$QCD in this way is somewhat serendipitous. We 
fit our chiral condensates to the 3-parameter form
\begin{equation}
\langle\bar{\psi}\psi(\beta)\rangle = b \langle M(a(\beta-\beta_c)+T_c)\rangle
\label{eqn:magfit}
\end{equation}
where $\langle\bar{\psi}\psi\rangle$ is the average of the magnitude of the
chiral condensate for $\chi$QCD and $\langle M \rangle$ is the magnetization of
the $O(2)$ spin model as a function of $T=1/J$. $T_c=1/0.454165$ is taken as
the critical temperature of the $O(2)$ spin model. $\beta_c$ the critical
value of $\beta=6/g^2$; $a$ and $b$ are the parameters of the fit. For the
$16^3 \times 8$ lattice condensate, the best fit is to the magnetization on a
$6^3$ lattice which had a $\chi^2/DOF=1.36$ with parameters $\beta_c=5.5358(4)$,
$a=23.7(8)$ and $b=0.380(4)$ over the range $5.50 \le \beta \le 5.57$. Fits
to other lattice sizes have $\chi^2/DOF=2.34$ ($4^3$), $\chi^2/DOF=1.58$
($5^3$), $\chi^2/DOF=1.55$ ($8^3$). For the $24^3 \times 8$ lattice, the best
fits to the condensate are to the magnetizations on lattice sizes $6^3$ which
has $\chi^2/DOF=1.85$ and parameters $\beta_c=5.5355(2)$, $a=53.5(1.3)$,
$b=0.312(2)$ and $8^3$ which has $\chi^2/DOF=1.87$ with $\beta_c=5.5360(2)$,
$a=30.7(6)$, $b=0.355(2)$ over the range $5.51 \le \beta \le 5.56$. Fits to
other lattice sizes have $\chi^2/DOF=2.63$ ($9^3$) and $\chi^2/DOF=3.16$ 
($5^3$). We are able to obtain reasonable fits over a range of lattice sizes,
since if our lattice QCD lattices were large enough to be fit by the
single scaling function of equation~\ref{eqn:fss}, we should be able to obtain
good fits of the above form for any $O(2)$ spin-model lattice large enough to
also be fit by a single finite-size-scaling function. Figure~\ref{fig:condfits}
shows the condensates on $16^3 \times 8$ and $24^3 \times 8$ lattices. The 
curves are the fits to the $O(2)$ spin models. We have also reanalyzed our
`data' from earlier simulations on an $18^3 \times 6$ lattice which we had
previously interpreted as showing tricritical scaling. We find that we can fit
these measurements to the magnetization for the $O(2)$ spin model on an $8^3$
lattice with $\chi^2/DOF=1.57$. Thus $N_t=6$ $\chi$QCD also shows consistency 
with $O(2)$ universality. These fits show the first evidence that the chiral
condensates are consistent with $O(2)$ scaling.

\begin{figure}[htb]
\epsfxsize=6in
\centerline{\epsffile{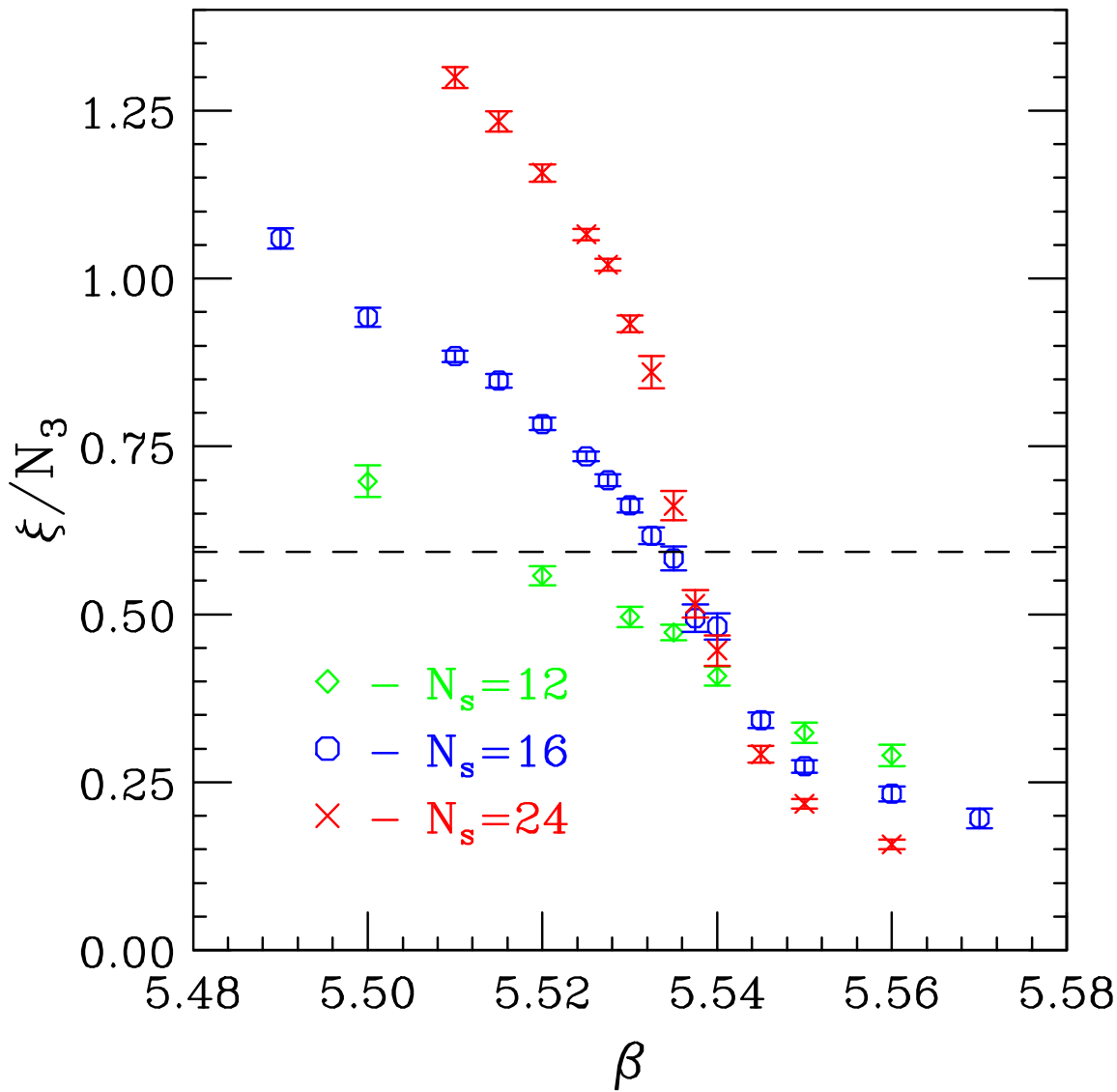}}
\caption{The second moment correlation lengths $\xi$ as functions of $\beta$
on $12^3 \times 8$, $16^3 \times 8$ and $24^3 \times 8$ lattices. Note that
$N_3=N_s$ since the spatial `box' is a cube in each case.}
\label{fig:xi}
\end{figure}

\begin{figure}[htb]
\epsfxsize=3.8in
\centerline{\epsffile{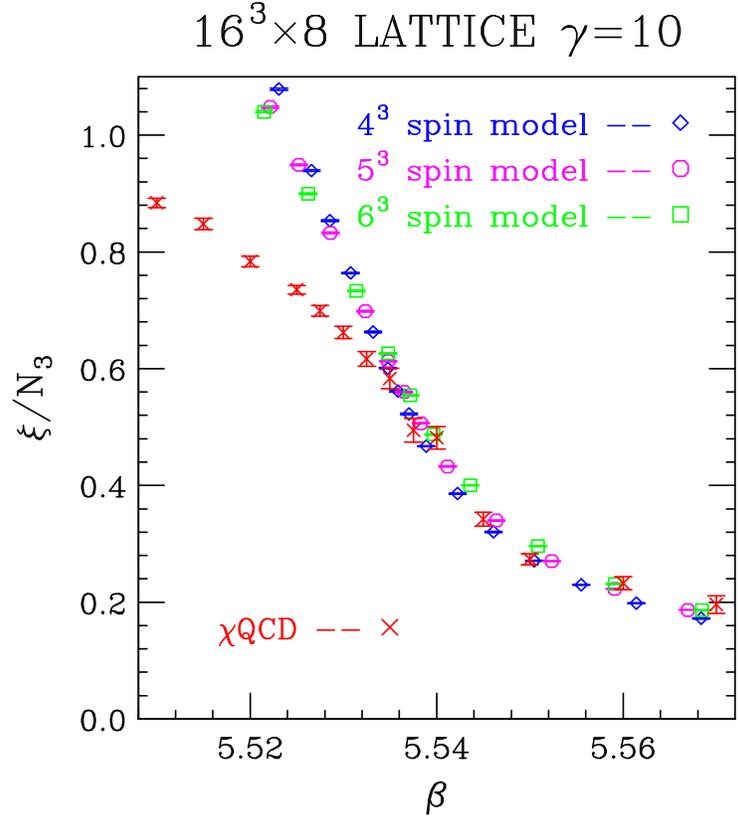}}
\vspace{0.25in}
\centerline{\epsffile{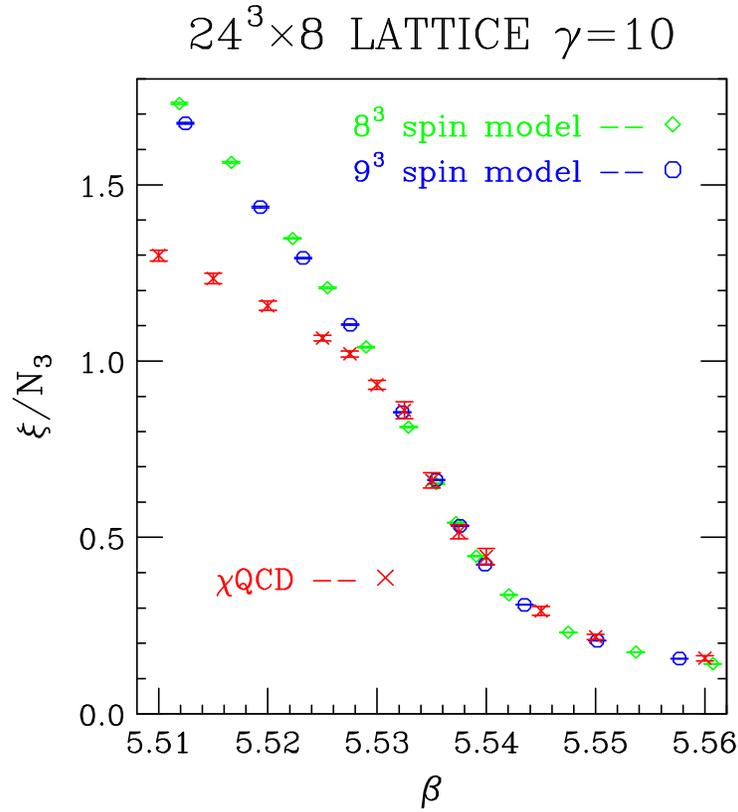}}
\caption{$\xi/N_3$ for $\chi$QCD and $\xi/L$ for the $O(2)$ spin model at
corresponding beta values: a) for the $16^3 \times 8$ lattice; b) for the
$24^3 \times 8$ lattice.}
\label{fig:xiscale}
\end{figure}

We next examine the correlation length $\xi$, defined as in 
equation~\ref{eqn:xi2} but with $\bm{\sigma}$ replacing $\bm{M}$. One 
observation that has been made in $O(2)$ spin models is that curves for $\xi/L$
($L$ is the box size) on different (large) lattice sizes cross at a point
where $\xi/L=0.593(2)$ \cite{Cucchieri:2002hu}, a value which is expected to
depend only on the universality class. Figure~\ref{fig:xi} shows $\xi$ for the
3 lattice sizes of our simulations. This gives some indication that the
intersection point of the correlation lengths is rising towards the universal
$O(2)$ value as the lattice size is increased. $\xi/N_3$ for lattice QCD and
$\xi/L$ for the spin model should define the same universal scaling function
in the high temperature phase (which is the only regime where the definition of
equation~\ref{eqn:xi2} has the interpretation as a correlation length), 
at corresponding values of $\beta$ and $T$, defined through
\begin{equation}
\beta=(T-T_c)/a+\beta_c,
\end{equation}
where $\beta_c$ and $a$ are obtained from the fits of equation~\ref{eqn:magfit},
except for $\langle\sigma\rangle$ rather than $\langle\bar{\psi}\psi\rangle$.
In figure~\ref{fig:xiscale} we have plotted the values of $\xi/N_3$ from our
$\chi$QCD simulations along with $\xi/L$ from the $O(2)$ lattice sizes which
show the best agreement. The agreement for the larger lattice is quite good in
the high temperature phase, that for the smaller lattice is only fair.

\begin{figure}[htb]
\epsfxsize=3.8in
\centerline{\epsffile{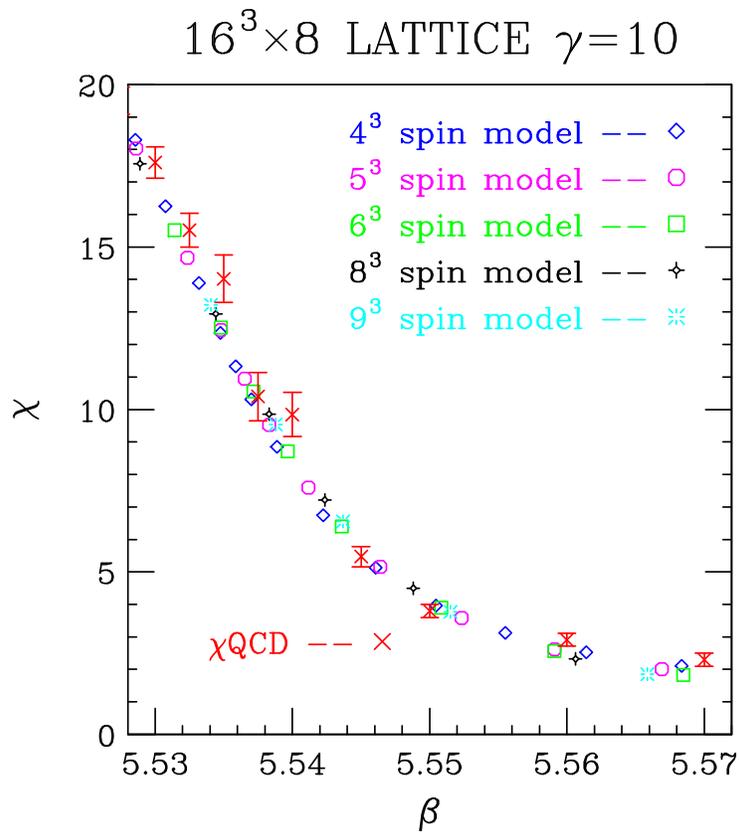}}
\vspace{0.25in}
\centerline{\epsffile{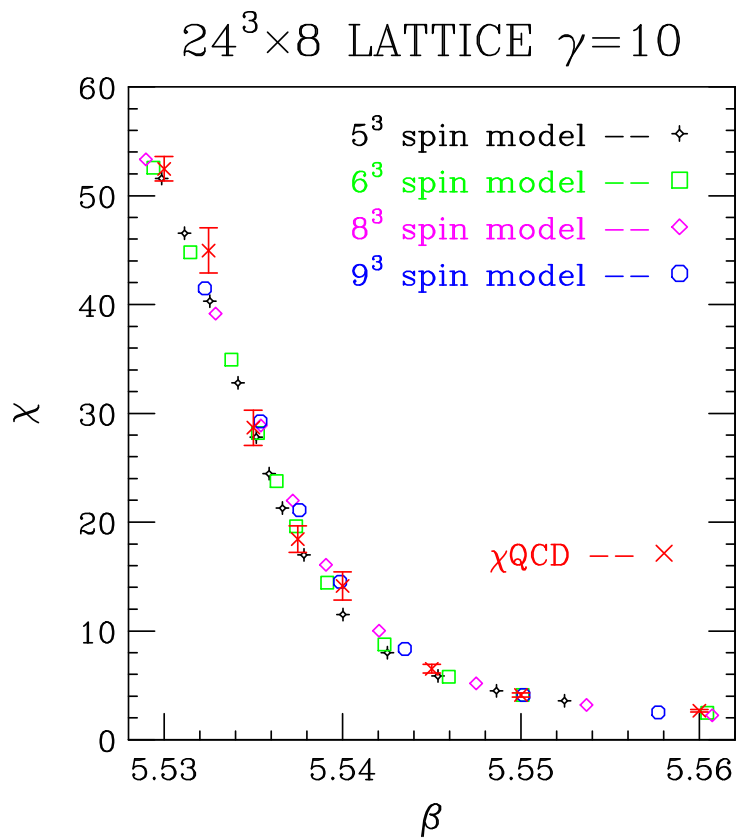}}
\caption{$\chi_\sigma$ for $\chi$QCD compared with the predictions from 
$\chi$ for the $O(2)$ spin model.}
\label{fig:chi}
\end{figure}

The final quantity which we wish to compare with the $O(2)$ spin model is the 
chiral susceptibility, which is the equivalent of that defined in 
equation~\ref{eqn:susc} with $\bm{M}$ replaced by $\bm{\sigma}$. As with the
correlation length discussed above, this really only has the interpretation
of susceptibility in the high temperature domain. The values of $a$ and $b$
from fits of $\langle\sigma\rangle$ using equation~\ref{eqn:magfit} give
predictions for the chiral susceptibility from the equivalent quantity for the
spin model
\begin{equation}
\chi_\sigma = V\langle\sigma^2+\pi^2\rangle = V b^2 \langle\bm{M}^2\rangle.
\end{equation}
We have plotted these predictions against our $\chi$QCD measurements in 
figure~\ref{fig:chi}. The agreement between the QCD `data' and spin-model
predictions is very good for the $24^3 \times 8$ lattice, less so for the
$16^3 \times 8$ lattice.

We have used Ferrenberg-Swendsen reweighting to obtain the peaks of the radial 
susceptibility for the $O(2)$ spin model as a function of lattice size, from
$4^3$ to $32^3$ lattices. We assume that the position of the peaks scales as
\begin{equation}
J_{peak}=J_c-K\,L^{-\theta}.
\end{equation}
This fit yields $J_c=0.45403(6)$ from a fit from lattice sizes from $6^3$ to
$32^3$ compared with the best estimate from the literature of 
$J_c=0.454165(4)$. The same fit yields $\theta=1.70(2)$ and $K=0.38(2)$. Hence
we conclude that this is a good method for estimating $J_c$ for the spin-model
and hence should be reasonable for estimating $\beta_c$ for $\chi$QCD.

We can now study the $O(2)$ spin model measurements to understand why it is
nearly impossible to interpret the critical behaviour of our lattice QCD
results directly without comparing them with the spin-model behaviour on small
lattices. In reference~\cite{Engels:2000xw} it was noted that it is only
possible to fit the magnetization curve in the infinite volume limit to the
leading order critical scaling form of equation~\ref{eqn:beta_m} over a very
small range of $|T_c-T|$ -- roughly 0--0.07. In terms of $J$ this means from
$J_c$ up to perhaps $0.47$. Above $J_c$, the leading finite-size corrections
to the magnetization are of order $1/L$. However, even when $J$ is as large as
$0.475$ extrapolating in $1/L$ only works for $L \gtrsim 12$. Hence, for
the $5^3$ to $9^3$ lattices which fit our $\chi$QCD `data', we
cannot perform the infinite volume extrapolation reliably. In the disordered
phase, the leading finite volume corrections are of order
$1/\sqrt{V}=1/L^{3/2}$. 
[This is true provided $L >> \xi$, since then one can divide the lattice into 
$N \propto V$ domains of extent $> \xi$. The orientations of the spins in these
domains will be independent of one another so that the magnetization of each
spin configuration will be ${\cal O}(\sqrt{N}/V)$ i.e. ${\cal O}(1/\sqrt{V})$.]
By the time $J$ is as large as $0.425$, departures
from this leading order extrapolation are already visible on the $9^3$
lattice. Hence simulations in the region of leading order scaling are
inaccessible to the small lattices of interest. Even if they were accessible,
they represent such a small region of $\beta$ (within $\approx 0.0024$ of the
transition for the $24^3 \times 8$ lattice) as to contain only 1 or 2 data
points. Reference~\cite{Engels:2000xw} finds good fits to the magnetization
over a much larger range using the scaling form
\begin{equation}
M = B(T_c-T)^{\beta_m}[1 + b_1(T_c-T)^{\omega\nu} + b_2(T_c-T)]
\end{equation}
-- this is the curve plotted in figure~\ref{fig:mag}. Even if we fixed all
3 critical exponents to their $O(2)$ values, this still has 4 parameters. By
the time we excluded enough points close to the transition and performed the
infinite volume extrapolations, such a fit for $\chi$QCD would have little
predictive power. As the authors of this paper and 
reference~\cite{Hasenbusch:1999cc} note, the narrowness of the scaling region
appears to be a property peculiar to the $O(2)$ spin model (at least to this
particular $O(2)$ spin model). [Note that the leading departure from scaling
(the term proportional to $b_1$) in this form corresponds to that in the
finite-size scaling form of equation~\ref{eqn:fss} with
$G(x) \rightarrow B b_1 T_c^{\omega\nu}x^{\omega\nu}$ as $x\rightarrow\infty$.
The final term is an order $L^{-1/\nu}$ correction term. Although, with the
parameters determined in \cite{Engels:2000xw} these 2 subleading terms are of
similar size, it is easy to convince oneself that if the second term is
omitted, a reasonable approximation to $M$ over the range of interest can be
obtained by readjusting parameters $B$ and $b_1$.]

\begin{figure}[htb]
\epsfxsize=4.0in
\centerline{\epsffile{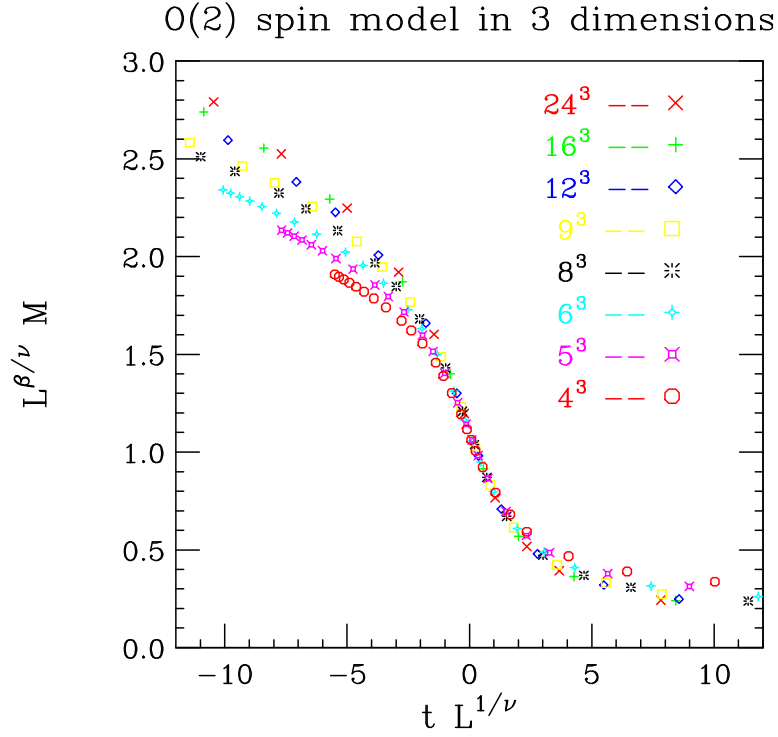}}
\vspace{0.25in}
\centerline{\epsffile{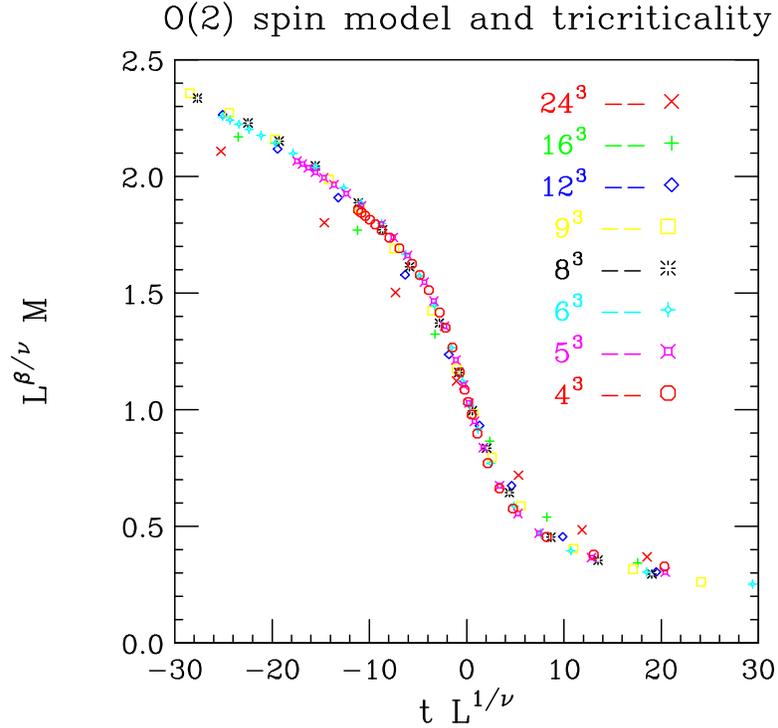}}
\caption{a) Finite size scaling analysis for the magnetization for the $O(2)$ 
spin model. b) Same as (a) but using tricritical exponents instead of $O(2)$
critical exponents.}
\label{fig:fss}
\end{figure}

Finite size scaling (with just the leading term of equation~\ref{eqn:fss})
apparently fails for the chiral condensates measured in our $\chi$QCD
simulations. It is therefore helpful to apply a finite size scaling analysis
to the magnetization for $O(2)$ spin models, including the data from small
lattices. To do this, we plot the scaling function $L^{\beta_m/\nu} M$ against
the scaling variable $t L^{1/\nu}$. Close to the transition ($t$ close to
zero), for large enough lattices, the data for all lattice sizes should follow
the same universal curve. Figure~\ref{fig:fss}a shows this curve using $O(2)$
critical indices $\beta_m$ and $\nu$. Near $t=0$ the curves for different
lattice sizes coincide. As one moves away from $t=0$ the curves start to
diverge. As the lattice sizes increase the curves follow one another for
larger ranges of $t L^{1/\nu}$, and it appears that they are approaching a
universal curve. As can be seen, the small lattices do not follow the
universal curve very far at all. While investigating such scaling, it is
interesting to try other values for $\beta_m$ and $\nu$. Our earlier, but now
suspect, identification of the $N_t=6$ transitions as being tricritical,
suggests we try the critical indices for a tricritical point, $\beta_m=0.25$
and $\nu=0.5$. The result is plotted in figure~\ref{fig:fss}b. The initial
impression is that this plot shows much better evidence for finite size
scaling than that with the correct critical exponents. A closer look reveals
that it is the larger lattices that show significant departures from the
`scaling curve'. This departure is small but significant for the $12^3$
lattice, larger for the $16^3$ lattice and embarrassingly large for the $24^3$
lattice. However, it is clear that if we only had the smaller lattices, we
would have concluded that the critical point was a tricritical point, not an
$O(2)$ critical point. Unfortunately this is essentially the situation for the
current round of lattice QCD simulations, so we need to be very careful.

\section{Discussion and conclusions}

We have simulated the finite temperature behaviour of lattice QCD with 2
flavours of staggered quarks using the $\chi$QCD action which allows us to run
at zero quark mass. Running with zero quark mass gives us direct access to the
second-order phase transition from hadronic matter to a quark-gluon plasma,
and allows us to to study its universality class. Attempts at determining the
nature of this transition for staggered quark actions from runs at finite mass
had led to a state of confusion, since the scaling behaviour near the
transition often appeared inconsistent with both the $O(4)$ universality class
expected for the continuum theory and the $O(2)$ universality class which one
would predict from considerations of the reduced symmetry of the staggered
fermions. The fermion masses that one is forced to use with standard staggered
actions are too large to clarify what is happening and some of the most
ambitious simulations of staggered fermion lattice QCD thermodynamics do not
even try to study the nature of the critical point. Notable exceptions include
a recent study of this transition at infinite coupling which shows clear
evidence for $O(2)$ universality \cite{Chandrasekharan:2003im}, and a study of
lattice QCD with 2 light adjoint quark flavours, where the chiral and
deconfinement transitions are distinct \cite{Engels:2005te}, which also shows
evidence for $O(2)$ universality of the chiral transition.

We present results for the finite temperature behaviour of 2-flavour lattice
QCD with the $\chi$QCD action on $12^3 \times 8$, $16^3 \times 8$ and 
$24^3 \times 8$ lattices. The decrease of the chiral condensate as the
critical point was approached from below appeared to be too abrupt for $O(2)$
or $O(4)$ universality. However, we found that it was possible to fit the
chiral condensates to the magnetization curves for an $O(2)$ spin model
on relatively small lattices $5^3$--$9^3$. This indicates that the transition
is consistent with $O(2)$ universality. The problems with more naive approaches
to measuring the details of the critical behaviour of this transition is that
they do not adequately account for finite volume effects close to the 
transition, and that for $O(2)$, the non-leading contributions to critical
scaling are important except so close to the transition as to be inaccessible
on the sizes of lattice we use. The spin model predictions for the correlation
lengths and chiral susceptibilities for lattice QCD are similar enough to our
measurements to add further support to this $O(2)$ interpretation. We also
made fits to the condensates from our $18^3 \times 6$ simulations, which we
had earlier interpreted as indicating tricritical behaviour, to the $O(2)$
spin model magnetizations on an $8^3$ lattice. These fits were good enough to
convince us that $N_t=6$ simulations also yield results consistent with $O(2)$
universality.

A finite size scaling analysis of our $O(2)$ spin model magnetizations gave
further clues as to why the behaviour of our lattice QCD measurements hinted
at a tricritical explanation. The approach of the $O(2)$ spin model 
magnetizations to the universal finite size scaling curve with increasing
lattice size is slow. However, if we perform a finite size scaling analysis
using the critical exponents of the tricritical point, rather than those
appropriate for $O(2)$, we find a rapid approach to an apparently universal
curve with increasing lattice size. It is only by going to larger lattices,
where the measurements start to diverge from this false scaling curve, that
the tricritical interpretation is invalidated. For lattice QCD we do not at
present have the luxury of being able to perform simulations on much larger
lattices.

The results we presented in this paper are for a single small fixed value (in
lattice units) of the 4-fermion coupling. It would be helpful if we could
perform simulations at another (even smaller) value of this coupling, since
the critical value of $\beta$, $5.535 \lesssim \beta_c \lesssim 5.536$ is
significantly above that for the standard staggered action at $N_t=8$, where
reference~\cite{Gottlieb:1996ae} estimates $\beta_c=5.44(3)$. We should also
extend our analyses to finite quark masses to study critical scaling involving
the critical exponent $\delta$. We note that we could extend our current
measurements to finite quark masses using Ferrenberg-Swendsen extrapolation.
However, our attempts to do such an extrapolation indicated that this is only
reliable for a range of masses where the condensate vanishes linearly with
decreasing quark mass. For finite mass studies one needs quark masses large
enough to get beyond this linear regime.

So far, we have only compared the critical behaviour of lattice QCD with the
$O(2)$ spin model. We need to extend our studies to spin models in other
universality classes as for example the $O(4)$ or the mean-field universality
classes, before we can say with any certainty whether the evidence for $O(2)$
universality is compelling. Such studies are planned for the near future,
since the spin-model simulations they involve are relatively inexpensive.

Finally, we need to address finite $dt$ effects. We have found that finite
$dt$ effects can change the apparent nature of the phase transition in our
studies of QCD at finite isospin chemical potential
\cite{Kogut:2005qg,Kogut:2005yu} (see also \cite{Philipsen:2005mj}). There the
problem was that the difference between the effective $\beta$, calculated by the
equipartition theorem from the molecular-dynamics kinetic energies and the
input $\beta$, a finite $dt$ effect, was much larger below the transition than
above, which distorted the transition. We have measured such $\beta$ shifts in
our $\chi$QCD simulations, and while they are as large as those measured in
our QCD at finite isospin simulations, they do not significantly change as we
cross the transition, and are therefore unlikely to distort this transition.
Nevertheless we intend to convert our simulations from hybrid
molecular-dynamics to exact RHMC simulations \cite{Clark:2003na}. Because the
lower bound of the spectrum of the $\chi$QCD Dirac operator is unknown, this
is not an entirely trivial conversion. However, our experience with using the
RHMC method to simulate QCD at finite isospin density, where the lower bound
on the Dirac operator is also unknown, indicates that RHMC simulations {\it can}
be applied in such cases \cite{Kogut:2006}.

\section*{Acknowledgements}

These $\chi$QCD simulations were performed on the IBM SP -- Seaborg at NERSC, 
the IBM SP -- Blue Horizon at SDSC/NPACI and the Tungsten cluster at NCSA. 
Access to the NSF machines was through an NRAC allocation. The spin model
simulations we performed on linux PCs in the HEP Division of Argonne National
Laboratory.

\end{document}